\documentclass[3p,twocolumn]{elsarticle}
\usepackage{amssymb}
\usepackage{graphicx}
\usepackage[breaklinks=true,colorlinks=true,linkcolor=blue,urlcolor=blue,citecolor=blue]{hyperref}

\begin{document}

\begin{frontmatter}

\title{Tuning of magnetic frustration in $S=1/2$ Kagom\'{e} lattices \{[Cu$_{3}$(CO$_{3}$)$_{2}$$(bpe)_{3}$](ClO$_{4}$)$_2$\}$_{n}$ and \{[Cu$_{3}$(CO$_{3}$)$_{2}$$(bpy)_{3}$](ClO$_{4}$)$_2$\}$_{n}$ through rigid and flexible ligands}

\author[iiserp]{M. O. Ajeesh}
\author[iiserp]{A. Yogi}
\author[iiserc]{M. Padmanabhan}
\author[iiserp]{R. Nath\corref{cor1}}
\ead{rnath@iisertvm.ac.in}
\cortext[cor1]{Corresponding author. Tel. : +91 0471 2599427}
\address[iiserp]{School of Physics, Indian Institute of Science Education and Research, Thiruvananthapuram-695016, India}
\address[iiserc]{School of Chemistry, Indian Institute of Science Education and Research, Thiruvananthapuram-695016, India}

\begin{abstract}
Single crystalline and polycrystalline samples of $S=1/2$ Kagom\'{e} lattices \{[Cu$_{3}$(CO$_{3}$)$_{2}$$(bpe)_{3}$](ClO$_{4}$)$_2$\}$_{n}$ and \{[Cu$_{3}$(CO$_{3}$)$_{2}$$(bpy)_{3}$](ClO$_{4}$)$_2$\}$_{n}$, respectively were synthesized. Their structural and magnetic properties were characterized by means of x-ray diffraction and magnetization measurements. Both compounds crystalize in a hexagonal structure (space group $P-6$) consisting of CuO$_4$ Kagom\'{e} layers in the $ab$-plane but linked along $c$ direction through either rigid $bpy$ or flexible $bpe$ ligands to form 3D frame works. Magnetic measurements reveal that both the compounds undergo ferromagnetic ordering ($T_{\rm C}$) at low temperatures and the $T_{\rm C}$ and the extent of frustration could be tuned by changing the nature of the pillar ligands. \{[Cu$_{3}$(CO$_{3}$)$_{2}$$(bpe)_{3}$](ClO$_{4}$)$_2$\}$_{n}$ which is made up of flexible $bpe$ ligands has a $T_{\rm C}$ of 5.7~K and a Curie-Weiss temperature ($\theta_{\rm CW}$) of -39.7 K giving rise to a frustration parameter of $\frac{|\theta_{\rm CW}|}{T_{\rm C}}$~$\simeq$~6.96. But the replacement of $bpe$ by a more rigid and electronically delocalized $bpy$ ligand leads to an enhanced $T_{\rm C}$~$\simeq$~9.3~K and a reduced frustration parameter of $\frac{|\theta_{\rm CW}|}{T_{\rm C}}$~$\simeq$~3.54.
%Using mean field approximation, the ratio of inter-layer to intra-layer exchange couplings was calculated for both the compounds.
\end{abstract}

\begin{keyword}
\noindent
 A. Magnetically ordered materials, D. Exchange and superexchange
\end{keyword}
\end{frontmatter}

\section{Introduction}
Frustrated magnetism is one of the most fascinating topics in condensed matter physics which attracts enormous attention from both experimental and theoretical researchers \cite{diep2005,mila2010}. The $S=1/2$ Kagom\'{e} lattices consisting of vertex-sharing triangular plaquettes of antiferromagnetically (AF) interacting $S=1/2$ ions in two-dimension (2D) are known to be highly frustrated in nature. In real materials, the combination of geometric frustration and strong quantum effects results in the persistence of spin fluctuations down to low temperatures. This leads to various unusual ground state properties including quantum spin liquid (QSL) \cite{balents2010}. The exact ground state of $S=1/2$ AF Kagom\'{e} lattice is still controversial since theory predicts different possibilities of the excitation spectra. It could be either gapped or gapless QSL ground state \cite{mila2010,yan2011,depenbrock2012,ran2007,hermele2008,iqbal2011}. With these variety of magnetic ground states, the Kagom\'{e} system becomes an interesting candidate with possibilities of novel low-energy and low-temperature magnetic properties.

Cu$_{3}$V$_{2}$O$_{7}$(OH)$_2$.2H$_2$O is the first material realization of $S=1/2$ AF Kagom\'{e} lattice in which the interaction was found to be anisotropic and it under goes a weak magnetic ordering at low temperature \cite{hiroi2001,yoshida2009}. Later, ZnCu$_3$(OH)$_6$Cl$_2$ was realized to be a perfect $S=1/2$ Kagom\'{e} lattice with isotropic AF Heisenberg interaction \cite{shores2005}. Neutron scattering and $\mu$SR experiments suggest that the material does not undergo magnetic ordering down to 50~mK \cite{helton2007,mendels2007}. Heat capacity follows a power law behaviour over a wide intermediate $T$-range \cite{helton2007} and NMR shift which is a measure of intrinsic susceptibility follows a power law. All these experimental features point to a possible gapless QSL ground state in ZnCu$_3$(OH)$_6$Cl$_2$. A gapless QSL ground state is also reported in vanadium oxyfluoride [NH$_4$]$_2$[C$_7$H$_{14}$N][V$_7$O$_6$F$_{18}$] \cite{clark2013}. Subsequently, a series of Kagom\'{e} lattice compounds have been synthesized where BaCu$_3$V$_2$O$_8$(OH)$_2$ was found to order at 9~K \cite{okamoto2009}, SrCr$_{9p}$Ga$_{12-9p}$O$_{19}$ shows a spin-glass transition at $T_{\rm g}$~$\simeq$~4~K, but retains fluctuations down to low temperature \cite{ramirez2000}, and also some of the compounds show different unusual ground state properties \cite{oba2005,rogado2002,rao2004}.

\begin{figure}
\begin{center}
\includegraphics [scale= .7]{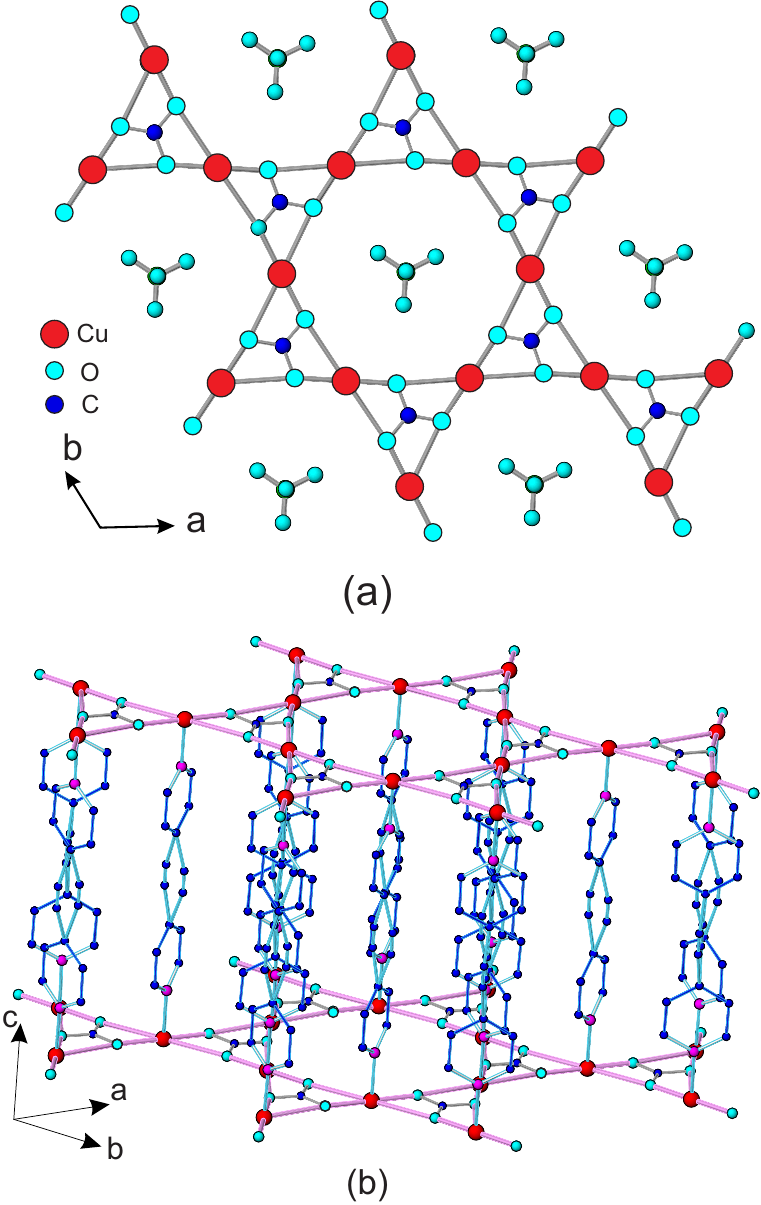}
\caption{\label{cubpeatom} (a) A section of the 2D Kagom\'{e} layer in Cu-$bpe$ formed by Cu$^{2+}$ and CO$_3$$^{2-}$ in the $ab$-plane. (b) 3D framework of Cu-$bpe$ showing the Kagom\'{e} layers pillared by the $bpe$ ligand along the $c$-axis.}
\end{center}
\end{figure}

Recently, \{[Cu$_{3}$(CO$_{3}$)$_{2}$$(bpe)_{3}$](ClO$_{4}$)$_2$\}$_{n}$ (abbreviated as Cu-$bpe$), with a planar S = 1/2 Kagom\'{e} layer pillared by a flexible organic ligand $bpe$ [$bpe$ = 1,2-di(4-pyridyl)ethane] was reported along with its structural and magnetic characterizations \cite{kanoo2009}. It has a hexagonal structure with $P-6$ space group and lattice constants $a$~=~9.3139(6)~\AA, $b$~=~9.3139(6)~\AA, and $c$~=~13.3133(9)~\AA. As seen from the crystal structure [Fig.~\ref{cubpeatom}(a)], the Cu$^{2+}$ ions are chelated to CO$_3$$^{2-}$ anions to form a perfectly 2D Kagom\'{e} layer in the $ab$-plane. The adjacent Kagom\'{e} layers are further interconnected via $bpe$ pillar ligands along the $c$-axis forming a three dimensional (3D) structure [Fig.~\ref{cubpeatom}(b)] \cite{kanoo2009}. In the layer the Cu - O bond lengths vary from 1.949(4) \AA\, to 2.763(5) \AA\, while both of the out of the plane Cu - N bond lengths are 1.987(3) \AA\, suggesting that the CuO$_4$N$_2$ octahdra takes a highly distorted structure. Magnetic susceptibility $\chi$($T$) shows a ferromagnetic transition ($T_{\rm C}$) at low temperature. The magnetization ($M$) vs. applied field ($H$) at $T$=2~K is reported to show a small hysteresis loop with a coercive field of 8.5~Oe. Heat capacity data do not show any well defined peak at the $T_{\rm C}$ \cite{kanoo2009}. On the other hand, $^{1}$H NMR spin-lattice relaxation rate ($1/T_1$) and spectral measurements by Kikuchi et al.,\cite{kikuchi2014} clearly suggest the ferromagnetic ordering at the $T_{\rm C}$.

In this paper we report synthesis, structural, and magnetic characterization of a new Kagom\'{e} system \{[Cu$_{3}$(CO$_{3}$)$_{2}$$(bpy)_{3}$](ClO$_{4}$)$_2$\}$_{n}$ [abbreviated as Cu-$bpy$] which contains a very rigid and electronically delocalized ligand $bpy$ [$bpy$ = 1,2-di(4-pyridyl)ethylene] interconnecting the adjacent layers. To understand the role of pillar ligands on the magnetic properties of the Kagom\'{e} systems we have also reinvestigated the recently reported Cu-$bpe$ Kagom\'{e} system with flexible $bpe$ pillar ligand having no electronic delocalization between the ends.

\section{Experimental Details }
The Kagom\'{e} compound Cu-$bpe$ was synthesized following the procedure reported in Ref.~\cite{kanoo2009}. Hexagonal plate-like violet crystals formed (90\% yield) were filtered and washed with distilled water and then air-dried. A good quality single crystal with dimension [$0.4\times 0.4~\times0.05$]~mm$^{3}$ was used for x-ray diffraction (XRD) (Bruker APEX-II machine with Mo K$_{\alpha 1}$ radiation of wavelength $\lambda$~=~0.71073~{\AA}) study. The phase purity of the sample was checked by powder XRD (PANalytical machine with Cu K$_{\alpha}$ radiation of $\lambda$~=~1.54060~{\AA} at room temperature) and matching with the simulated pattern.

Since the above procedure failed to produce the expected Kagom\'{e} compound with the $bpy$ ligand, we made use of a modified method for the synthesis of new compound Cu-$bpy$. To a solution of Cu(ClO$_{4}$)$_{2}$.6H$_{2}$O (1.853~g, 5~mM, in 100~mL distilled water and 50~mL aqueous 20\% ammonia), a $bpy$ solution (0.912~g, 5~mM, in 50~mL aqueous 20\% ammonia and 20~mL methanol) was added and stirred for 1 hour. After 3 days, blue-violet polycrystalline powder of Cu-$bpy$ was formed in high yield (92\%) by fixing atmospheric CO$_2$ as CO$_3$$^{2-}$ anion. The solid compound was filtered, washed with water and methanol and then air-dried. Our repeated attempts to get single crystal form of Cu-$bpy$ by varying experimental conditions turned out to be unsuccessful. However, the polycrystalline sample obtained was phase-pure form of the Kagom\'{e} structure as confirmed by powder XRD studies along with FTIR and CHN analysis.

DC magnetic susceptibility $\chi$ (= $M/H$, where $M$ is magnetization and $H$ is applied magnetic field) was measured as a function of $T$~(2 K $\leq$ $T$ $\leq$ 300 K) and at different applied fields. A magnetization isotherm ($M$ vs. $H$) was measured at $T$ = 4 K by varying the magnetic field up to 7 T. Both $\chi$ and magnetization isotherm were measured using a 7~T SQUID - VSM (Quantum Design).
AC magnetic susceptibility ($\chi_{ac}$ = $\chi'$ + i$\chi''$, where $\chi'$ is the real part and $\chi''$ is the imaginary part of the susceptibility) measurements were performed using a separate AC susceptibility setup (Cryobind AC susceptometer) as a function of $T$ and at different frequencies.

\section{Results}
\subsection{X-ray diffraction}
\begin{figure}
\begin{center}
\includegraphics [scale=0.63]{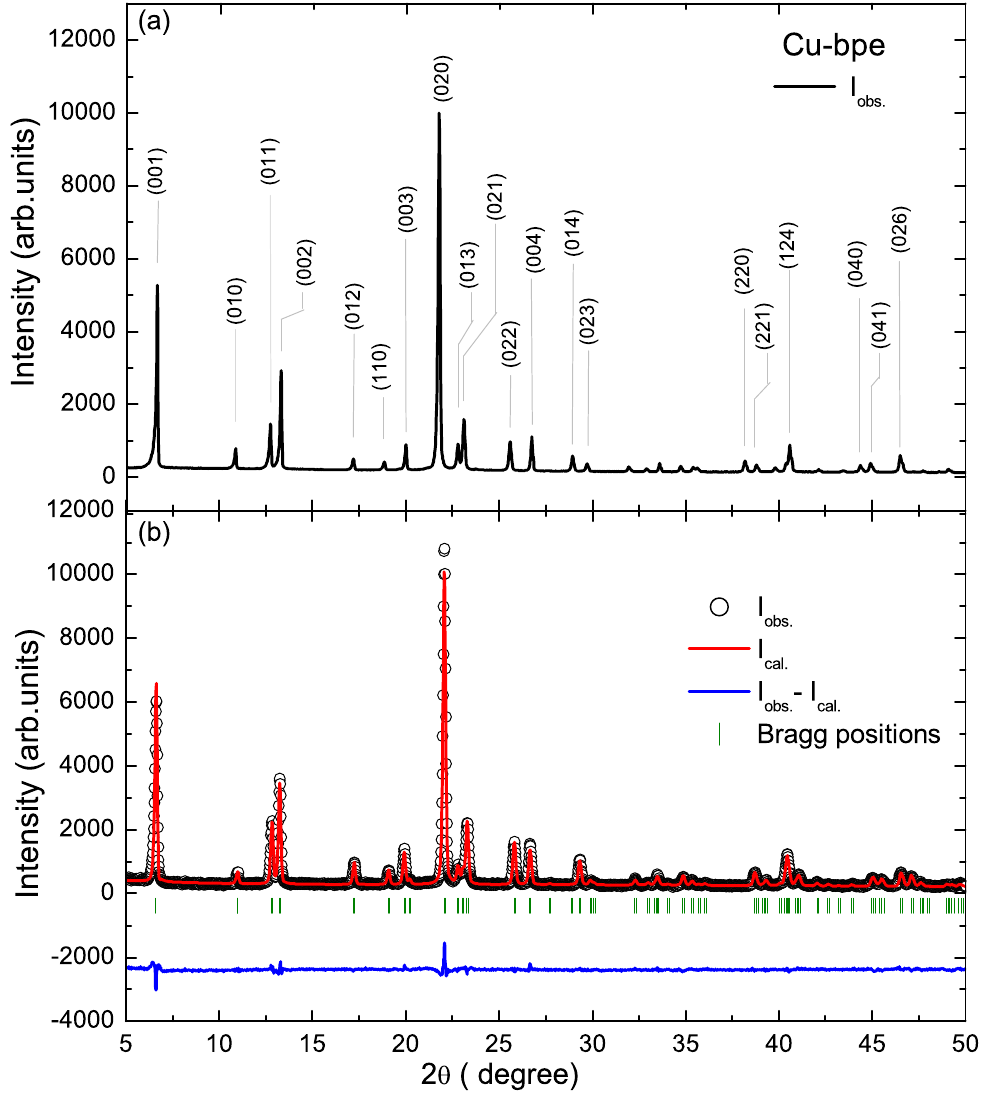}
\setlength{\abovecaptionskip}{5pt}
\caption{\label{pxrd}(a) Powder XRD pattern of Cu-$bpe$ recorded at room temperature. ($hkl$) values of the corresponding peaks are indicated. (b) Powder XRD pattern (open circles) recorded at room temperature for Cu-$bpy$. The solid line through the experimental points are the Rietveld refinement profile calculated using structural parameters of Cu-$bpe$. The short vertical bars mark the fitted Bragg peak positions. The lowermost curve represents the difference between the experimental and calculated intensities.}
\end{center}
\end{figure}

Single crystal XRD was performed on a good quality single crystal of Cu-$bpe$. It is found to be crystallizing in the hexagonal structure with space group $P-6$ and lattice constants $a$~=~9.4102(6)~\AA, $b$~=~9.4102(6)~\AA, $c$~=~13.3253(8)~\AA, and V$_{\rm cell}$~=~1021.89(11)~\AA$^3$ which are consistent with the previous report \cite{kanoo2009}. Powder XRD data of Cu-$bpe$ is shown in Fig.~\ref{pxrd}(a) where all the peaks are indexed based on the above structural data. As shown in Fig.~\ref{pxrd}(b), very similar peaks appear in the powder XRD pattern of Cu-$bpy$ suggesting that Cu-$bpy$ is isostructural to Cu-$bpe$. Rietveld refinement was performed using the FullProf software with the structural parameters reported in Ref.~\cite{kanoo2009} as the starting parameters \cite{carvajal1993}. Since $bpy$ is restricted in the $trans$-configuration because of the -HC=CH- double bond between the pyridyl moieties, C and H atoms corresponding to the orientational disorder (C9 and H9) seen in Cu-$bpe$ were removed and the occupancy of C8 and H8 atoms were corrected to one. The room-temperature powder XRD pattern along with the Rietveld refinement fit is shown in Fig.~\ref{pxrd}(b). The Rietveld refinements confirmed the phase purity of Cu-$bpy$ and also the hexagonal structure (space group: $P-6$), same as that of Cu-$bpe$. The best fit was obtained with $\chi^2$~$\simeq$~1.86 and the lattice parameters obtained from refinement are $a$~=~9.3024(5)~\AA, $b$~=~9.3024(5)~\AA, $c$~=~13.3724(7)~\AA, and V$_{\rm cell}$~=~1157.17~\AA$^3$.

\subsection{DC Magnetization}
\begin{figure}
\begin{center}
\includegraphics [scale=.7]{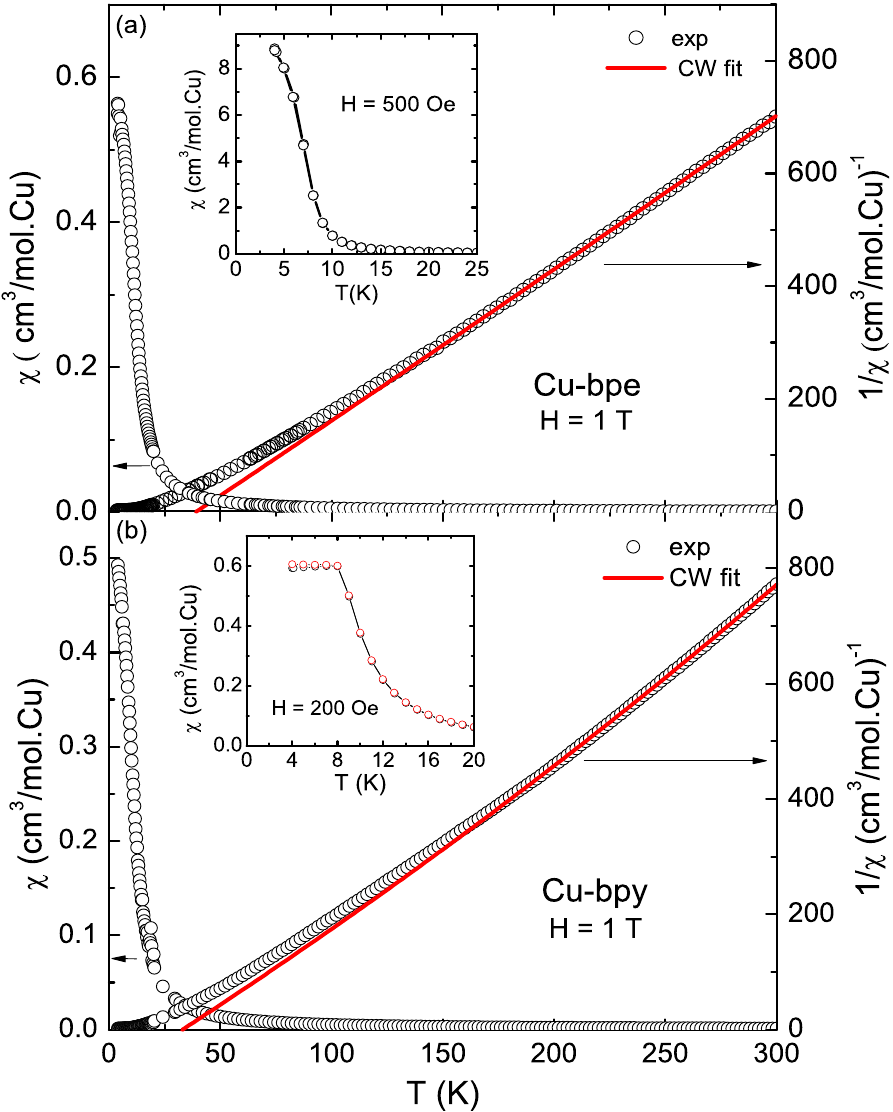}
\setlength{\abovecaptionskip}{5pt}
\caption{\label{Chidc}DC Magnetic susceptibility $\chi$ (left $y$-axis) and $\chi^{-1}$ (right $y$-axis) of (a) Cu-$bpe$ and (b) Cu-$bpy$ vs. temperature $T$ at an applied magnetic field $H$ = 1 T. The solid red lines are the CW fits to the data for the temperature range 189~-~300 K and 154~-~185 K for Cu-$bpe$ and Cu-$bpy$, respectively. Insets: ZFC and FC susceptibilities of Cu-$bpe$ and Cu-$bpy$ vs. $T$ at low fields.}
\end{center}
\end{figure}

The DC magnetic susceptibility ($\chi$) data of Cu-$bpe$ and Cu-$bpy$ as a function of $T$ at an applied magnetic field of 1~T are shown in Fig.~\ref{Chidc}(a) and \ref{Chidc}(b), respectively. At high $T$s, $\chi$($T$) increases in a Curie-Weiss manner with decreasing $T$ and then shows a rapid increase at low temperature. This rapid increase which occurs below about 20~K for both the compounds is an indication of ferromagnetic (FM) correlation.

In order to extract the magnetic parameters, the $\chi$($T$) data at high $T$ region are fitted by the expression
\begin{equation}\label{CW fit}
\chi=\chi_{0}+\frac{C}{T+\theta_{\rm CW}}\,,
\end{equation}
%\newline
where $\chi_{0}$ is the temperature independent contribution consisting of core diamagnetism of the core shells and Van-Vleck paramagnetism of the open shells of Cu$^{2+}$ ions. The second term is the Curie-Weiss (CW) law with $C$ (= $N_A\mu_{\rm eff}^{2}/3k_B$, where $N_{A}$ is Avogadro's number, $\mu$$_{\rm eff}$ is the effective magnetic moment, and $k_{B}$  is Boltzmann's constant) being the Curie constant and $\theta_{\rm CW}$ the Curie-Weiss temperature. The extracted parameters for both the compounds are mentioned in Table-\ref{table1}. The calculated values of $\mu$$_{\rm eff}$ from $C$ for both the compounds are in good agreement with the theoretical value [$\mu$$_{\rm eff}$ = $ g \sqrt{S(S+1)}$$\mu$$_B$] of 1.73~$\mu$$_B$ expected for $S=1/2$ systems assuming $g$-factor $g$~=~2. The core diamagnetic susceptibility ($\chi_{\rm {core}}$) was calculated by adding the contributions from individual ions to be $-5.2534\times 10^{-4}$~cm$^{3}$/mol and $-4.9126\times 10^{-4}$~cm$^{3}$/mol for Cu-bpe and Cu-bpy, respectively \cite{magnetochemistry}. The Van-Vleck paramagnetic susceptibility ($\chi_{\rm VV}$) was estimated by subtracting $\chi_{\rm {core}}$ from $\chi_{0}$ to be $\sim 4.974\times 10^{-4}$~cm$^{3}$/mol and $\sim 3.012\times 10^{-4}$~cm$^{3}$/mol for Cu-bpe and Cu-bpy, respectively.

\begin{@twocolumnfalse}
\begin{table*}
\scriptsize
\begin{center}
\setlength{\belowcaptionskip}{15pt}
 \caption{Magnetic parameters obtained by fitting Eq.~\ref{CW fit} to $\chi$($T$) data in the $T$-range (189~-~300 K) and (154~-~185 K) for Cu-$bpe$ and Cu-$bpy$, respectively. $\mu$$_{\rm eff}$ was calculated using the experimental $C$ value.}
\label{table1}
\begin{tabular}{|c|c|c|c|c|}
  \hline  \hline
          & $\chi_{0}$   & $C$  & $\theta_{\rm CW}$  & $\mu$$_{\rm eff}$  \\
          & (cm$^{3}$/mol.Cu) & (cm$^3$.K/mol.Cu) & (K) & ($\mu$$_B$) \\
\hline
Cu-$bpe$ & -0.279(5)$\times$10$^{-4}$ & 0.378(4) & -39.7(9) & 1.738(2) \\
Cu-$bpy$ & -1.904(4)$\times$10$^{-4}$ & 0.397(6) & -33.0(2) & 1.781(3) \\
\hline  \hline
\end{tabular}
\end{center}
\end{table*}
\end{@twocolumnfalse}

\begin{figure}
\begin{center}
\includegraphics [scale=.85]{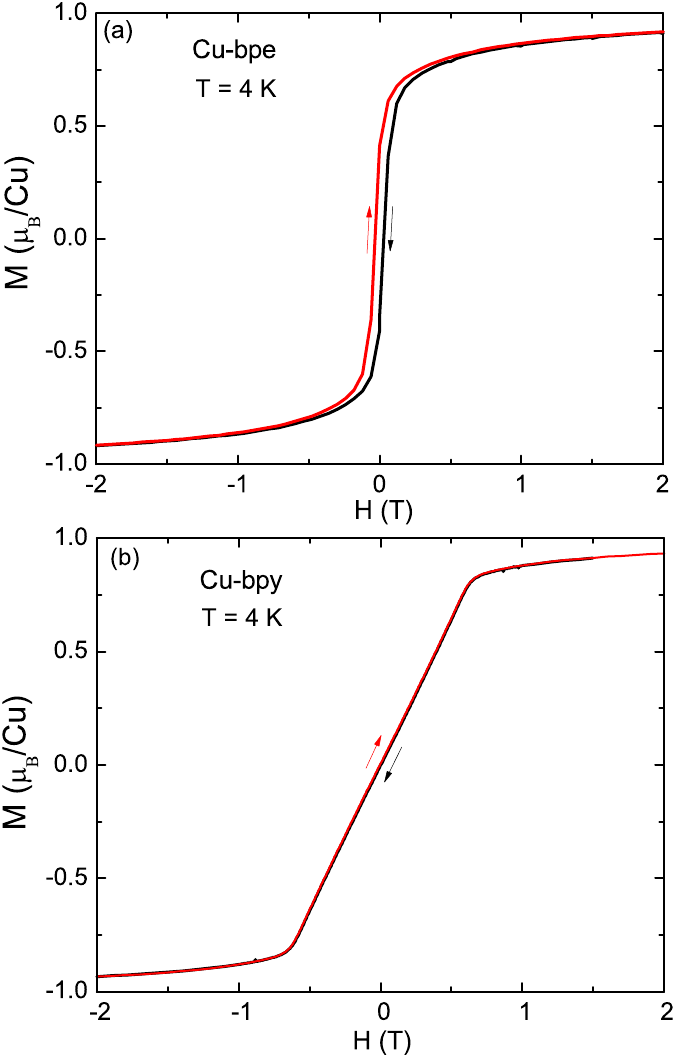}
\setlength{\abovecaptionskip}{5pt}
\caption{\label{MvH} $M$ vs. $H$ for (a) Cu-$bpe$ and (b) Cu-$bpy$ at $T$ = 4~K.}
\end{center}
\end{figure}

We also measured the low temperature Zero Field Cooled(ZFC) and Field Cooled(FC) susceptibilities at a small applied field [see insets of Fig.~\ref{Chidc}(a) and \ref{Chidc}(b)]. No significant splitting between ZFC and FC susceptibilities was observed for both the compounds. This confirms the absence of any glassy component and spin reorientation effect in the ordered state. To further confirm the FM correlation and also to check whether there is any field induced effect at low temperature, we measured $M$ vs. $H$ at $T$~=~4 K for both the compounds (Fig.~\ref{MvH}). A typical hysteresis loop was observed with a coercive field $H_{\rm c}$~$\simeq$~600~Oe and 60~Oe for Cu-$bpe$ and Cu-$bpy$, respectively suggesting a weak ferromagnetism in both the compounds below $T_{\rm C}$. Moreover, the value of $H_{\rm c}$ reported in Ref.~\cite{kanoo2009} for Cu-$bpe$ is two orders of magnitude less than our observed value. The magnetization was found to reach a saturation value of 0.9 $\mu$$_B$/Cu at about 1.6~T for both the compounds which is close to the expected spin only value of (=~$gS$$\mu$$_B$) 1 $\mu$$_B$, assuming $g$~=~2. No signature of any field induced transition was obtained in the $M$ vs. $H$ curve.

\begin{figure}
\begin{center}
\includegraphics [scale=1]{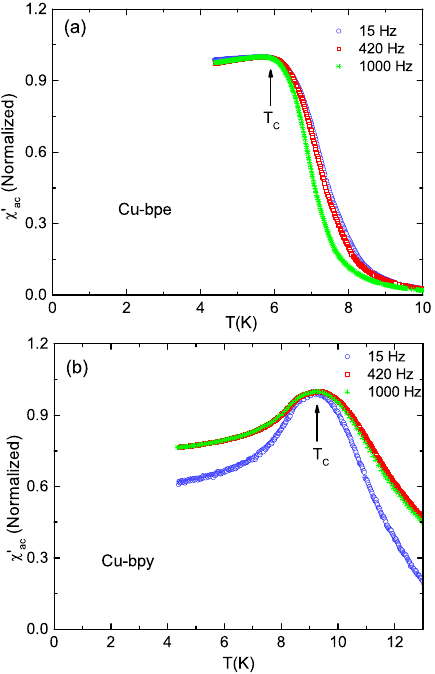}
\setlength{\abovecaptionskip}{5pt}
\caption{\label{chiac} Real part of AC magnetic susceptibility ($\chi_{ac}'$) of (a) Cu-$bpe$ and (b) Cu-$bpy$ as a function of $T$ at an applied AC field of $H_{\rm ac}$~=~0.177~Oe for different frequencies.}
\end{center}
\end{figure}
\subsection{AC Susceptibility}
Since AC susceptibility ($\chi_{ac}$) is very sensitive to magnetic ordering, one can gain information about the nature of the magnetic ordering by measuring the frequency dependent $\chi_{ac}$. Figure~\ref{chiac}(a) and \ref{chiac}(b) present the real part of AC susceptibility ($\chi_{ac}'$) of Cu-$bpe$ and Cu-$bpy$, respectively as a function of $T$ measured at an applied AC field of $H_{\rm ac}$ = 0.177~Oe for three different frequencies 15 Hz, 420 Hz, and 1000 Hz. The $\chi_{ac}'$ for Cu-$bpe$ shows a broad maximum at $T_{\rm C}$~$\simeq$~5.7~K while for Cu-$bpy$ the maximum is more pronounced at $T_{\rm C}$~$\simeq$~9.3 K, reflecting their $T_{\rm C}$s. The broad maxima for both the compounds remain frequency independent, indicating the absence of spin-glass behaviour below $T_{\rm C}$ \cite{mydosh1996}. This is also a characteristic feature of bulk ferromagnet.

\section{Discussion and Conclusion}
In the crystal structure,  the angle $\angle${Cu-O-Cu} is 170$^0$ in the Kagom\'{e} plane. According to Goodenough-Kanamori rule, the superexchange interaction through a non-magnetic ion will favour ferromagnetism if the angle between them is close to 90$^0$ and antiferromagnetism if the angle is close to 180$^0$ \cite{goodenough1963,kanamori1959}. This suggests that the intra-layer exchange interaction ($J_1$) in both the compounds should be antiferromagnetic.
According to the mean field theory $\theta_{\rm CW}$ is the sum of all the exchange couplings \cite{misenheimer1961}. But the negative value of $\theta_{\rm CW}$ indicates that the dominant interaction is FM. Since $J_1$ is expected to be AFM, this dominant FM interaction is most likely originating from the inter-layer superexchange ($J_2$) through $bpe$ and $bpy$ ligands between Cu$^{2+}$ ions in the adjacent layers. In the present case, estimation of individual exchange couplings from the values of $\theta_{\rm CW}$ and $T_C$ is not trivial since it may involve another exchange coupling $J_3$ along the hexagon. An estimation of individual exchange couplings through electronic band structure calculations would be useful to understand the exchange mechanism in such compounds \cite{nath2014}. FM inter-layer interactions are also reported in Kagom\'{e} systems Cu(1,3-bdc) and KFe$_3$(OH)$_6$(SO$_4$)$_2$  with similar geometries \cite{takano1970,nytko2008}. There are also other copper based metal-organic complexes having similar ligands which are reported to show dominant FM interactions \cite{suman04}.

As discussed before, the distorted CuO$_4$N$_2$ octahedra gives rise to an anisotropic CuO$_4$ arrangement in the plane. In such a geometry, the d$_{x^2-y^2}$ orbital of Cu$^{2+}$ ion and 2p orbital of O$^{2-}$ ion, which are responsible for super-exchange are aligned orthogonal to each other \cite{jeffrey75}. Due to this kind of orbital arrangements, the super-exchange between Cu$^{2+}$ ions are very weak in the plane and the interaction between the planes via ligands dominates.

%The relative strength of the intra-layer ($J_1$) and inter-layer ($J_2$) exchange interactions can be estimated from mean field approximation %\cite{yogesh2009}. Assuming that a spin interacts with same number of spins both within the same layer and with adjacent layers, one can apply the %mean field treatment for two-sublattice model to get

%\begin{equation}\label{meanfield}
%\frac{T_{\rm C}}{\theta_{\rm CW}} = \frac{\lambda_1-\lambda_2}{\lambda_1+\lambda_2} = \frac{J_1-J_2}{J_1+J_2}\,,
%\end{equation}
%where, $\lambda_1$ and $\lambda_2$ are the mean field coupling constants for spins in the same layer and adjacent layers, respectively. By %substituting the experimental values of $T_{\rm C}/\theta_{\rm CW}$ in Eq.~\ref{meanfield}, we find $J_2/J_1$ to be $-$1.33 and $-$1.8 for Cu-$bpe$ %and Cu-$bpy$, respectively. This confirms that, $J_2$ is stronger than $J_1$ and hence dominates the ordering at low temperature.

\begin{figure}
\begin{center}
\includegraphics [scale=.6]{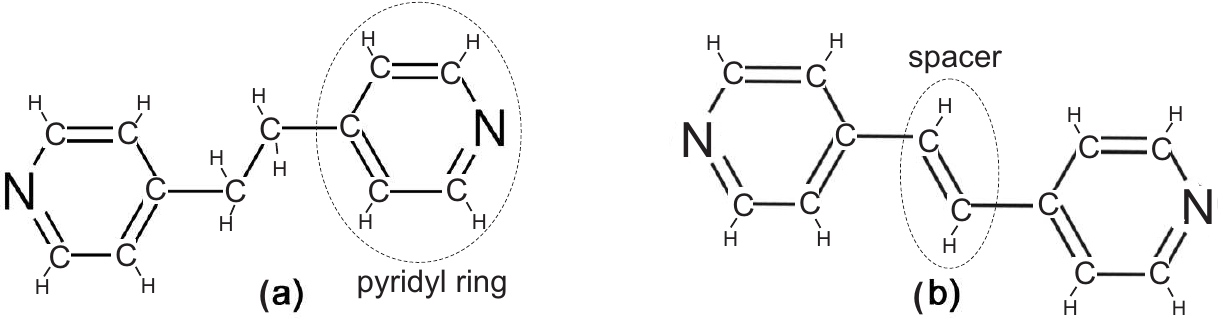}
\setlength{\abovecaptionskip}{5pt}
\caption{\label{ligands} Structure of (a) $bpe$ and (b) $bpy$ ligands. The portion enclosed by the circle in (a) is pyridyl ring and in (b) is the spacer.}
\end{center}
\end{figure}

With the estimated values of $\theta_{\rm CW}$ and $T_{\rm C}$, the frustration in the spin system can be quantified. The frustration parameter $f$, which is an empirical measure of frustration \cite{ramirez1994} is given by,
\begin{equation}\label{fpara}
f = \frac{|\theta_{\rm CW}|}{T_{\rm C}}\,.
\end{equation}
Typically spin systems with $f$ $>$ 5-10 are considered to be moderately frustrated. The frustration parameter $f$ for Cu-$bpe$ and Cu-$bpy$ was found to be 6.96 and 3.54, respectively suggesting that Cu-$bpe$ is more frustrated than Cu-$bpy$. This can be explained by the structural differences seen in $bpe$ and $bpy$ ligands which is evident from Fig.~\ref{ligands}. The difference between the $bpe$ and $bpy$ ligands is that, the spacer between the pyridyl rings is having a single bond ( -CH$_2$-CH$_2$- ) in $bpe$ compared to a double bond ( -HC=CH- ) in $bpy$. The presence of double bond in Cu-$bpy$ is responsible for facilitating strong exchange interaction between the Kagom\'{e} layers compared to Cu-$bpe$. Therefore, replacement of the flexible $bpe$ by a more rigid and electronically delocalized $bpy$ ligand leads to an increase of $T_{\rm C}$ from 5.7~K to 9.3~K and decrease of $f$ from 6.96 to 3.54.

Ground state properties of such complexes can further be tuned by modifying the nature of the pillar ligand and increasing the inter-planar distance. Owing to the perfectly planar Kagom\'{e} structure and the tunability of inter-layer exchange coupling through appropriate linker molecules, this kind of Kagom\'{e} systems would serve as promising candidates for exploring exotic ground state properties.

\section*{Acknowledgments}
We would like to thank Dr. B. R. Sekhar (IOP Bhubaneswar) for SQUID-VSM and Dr. P. S. Anil Kumar (IISc Bangalore) for AC susceptibility measurements. DST India is acknowledged for financial support.

\end{document}